
\catcode`@=11

\def\pri{^\prime}


\def\titlepage{\FRONTPAGE\paperstyle\ifPhysRev\PH@SR@V\fi
   \ifp@bblock\p@bblock\fi}

\def\p@bblock{\begingroup \tabskip=\hsize minus \hsize
   \baselineskip=1.5\ht\strutbox \topspace-2\baselineskip
   \halign to\hsize{\strut ##\hfil\tabskip=0pt\crcr
   \the\Pubnum\cr \the\date\cr \the\pf\cr \the\pubmemo\cr}\endgroup}

\Pubnum={KUCP-\the\pubnum}
\pubnum={00}
\date={\monthname,\ \number\year}
\newtoks\pf
\pf={T/TE/AS}
\def\datenum{\number\month .\number\day}
\newtoks\pubmemo
\pubmemo={\it Ver. \datenum}

\def\abstract{\vskip\frontpageskip\centerline{\twelverm ABSTRACT}
              \vskip\headskip }
\def\author#1{\vskip 0.3cm \titlestyle{\twelvecp #1}\nobreak}

\def\kyoto{\centerline{\sl Department of Physics}
          \centerline{\sl College of Liberal Arts and Sciences}
          \centerline{\sl Kyoto University}
          \centerline{\sl Yoshida, Kyoto 606, Japan}}


\def\NP{Nucl.~Phys.~}
\def\PR{Phys.~Rev.~}
\def\PRL{Phys.~Rev.~Lett.~}
\def\PL{Phys.~Lett.~}

\date={\monthname \ \number\day,  \number\year}

   \newdimen\squareht
   \newdimen\squarewd

   \newbox\squarebox

   \def\boxit#1{\vbox{\hrule\hbox{\vrule\kern3pt%
       \vbox{\kern3pt#1\kern3pt}\kern3pt\vrule}\hrule}%
   }

   \def\square{%
      \setbox\squarebox=\boxit{\hbox{\phantom{x}}}
      \squareht = 1\ht\squarebox
      \squarewd = 1\wd\squarebox
      \vbox to 0pt{
          \offinterlineskip \kern -.9\squareht
          \hbox{\copy\squarebox \vrule width .2\squarewd height .8\squareht
              depth 0pt \hfill
          }
          \hbox{\kern .2\squarewd\vbox{\hrule height .2\squarewd
          width \squarewd}
          }
          \vss
      }
   }


\catcode`@=12

\catcode`\@=11 
%
%
\newskip\frontpageskip
\newtoks\Pubnum   \let\pubnum=\Pubnum
\newtoks\Pubnumn  
\newtoks\Pubtype  \let\pubtype=\Pubtype
\newif\ifp@bblock  \p@bblocktrue
\def\titlepage{\FRONTPAGE\papers\ifPhysRev\PH@SR@V\fi
   \ifp@bblock\p@bblock \else\hrule height\z@ \rel@x \fi }
\def\endpage{\vfil\break}
\frontpageskip=12pt plus .5fil minus 2pt
\Pubtype={}
\Pubnum={}
\Pubnumn={KUCP-??}
\def\p@bblock{\begingroup \tabskip=\hsize minus \hsize
   \baselineskip=1.5\ht\strutbox \topspace-2\baselineskip
   \halign to\hsize{\strut ##\hfil\tabskip=0pt\crcr
       \the\Pubnum\crcr
       \the\date\crcr\the\pubtype\crcr}\endgroup}
\def\abstract{\par\dimen@=\prevdepth \hrule height\z@ \prevdepth=\dimen@
   \vskip\frontpageskip\centerline{ABSTRACT}\vskip\headskip }
\catcode`\@=12 
\def\kyoto{
{\sl
\centerline{Department of Fundamental Sciences,
     Faculty of Integrated Human Studies}
\centerline{Kyoto University, Yoshida, Kyoto 606-01, Japan}}}
%

%

      \def\kk{{k \kern-2.05mm k \kern-2.05mm k}}
      \def\bfk{{k \kern-1.83mm k }}
   \def\NN{{N \kern-4.50mm N \kern-4.50mm N}}

 \pubnum{KUCP-64}
 \date{January 1994}
 \pubtype{}

\titlepage
\vskip 1cm
 \title{\bf Representation Blocks of Conformal Fields for
    the  $\NN$=4  SU(2)$_\kk$ Superconformal Algebras}

\vskip 5mm
      \author{Satoshi Matsuda\foot{Work supported in part by
         the Grant-in-Aid for Scientific Research
         on Priority Area $\lq\lq$Infinite Analysis''
         ($\#$05230029) from the Ministry
         of Education, Science and Culture.}}
\vskip 0.3cm
 \kyoto

\vskip 1cm

 \abstract

The representation theories of the
SU(2)$_k$-extended $N$=4 superconformal algebras (SCAs)
with $arbitrary$ level $k$
are developed being based on their Feigin-Fuchs representations
found recently by the present author.
A basic unit of the representation blocks consisting
of eight \lq\lq boson-like\rq\rq\
and eight \lq\lq fermion-like\rq\rq\
conformal fields is found to describe arbitrary representations
of the $N$=4 SU(2)$_k$ SCAs,
including {\it unitary} and {\it nonunitary} representations.
The transformation properties of the fundamental sets of the
conformal fields
under the $N$=4 SU(2)$_k$ superconformal symmetries
are given.
Then, the whole sets of the charge-screening operators
of the $N$=4 SU(2)$_k$ SCAs
are identified out of the sixteen conformal fields
in the basic unit of the representation blocks.
The conditions for the {\it eligible} charge-screening operators
are  analyzed in terms of the continuous parameters
which enter in our vertex-operator forms for
the fundamental conformal fields of the representation blocks.

\vskip 1cm
\vskip .8cm
\endpage

\def\NP{ Nucl. Phys.}
\def\PR{ Phys. Rev.}
\def\PRL{ Phys. Rev. Lett.}
\def\PL{ Phys. Lett. }
\def\PTP{ Progr. Theor. Phys.}

\def\MPL{ Mod. Phys. Lett.}


\sequentialequations


\normalspace

\PHYSREV




\REF\yu{M.~Yu \journal\PL &196B (87) 345;
{\sl Nucl. Phys.} {\bf B294} (1987), 890.}

\REF\eguchi{T.~Eguchi and A.~Taormina \journal\PL &196B (87) 75;
{\bf B200} (1988), 315.}
\REF\taormina{T.~Eguchi and A.~Taormina \journal\PL &210B (88) 125.}

\REF\petersen{J.~L.~Petersen and A.~Taormina \journal\NP &B331 (1990) 556;
{\bf B333} (1990), 833.}

\REF\musf{S.~Matsuda and T.~Uematsu \journal\PL &220B (89) 413.}
\REF\musd{S.~Matsuda and T.~Uematsu \journal\MPL &A5 (90) 841.}

\REF\kent{A.~Kent and H.~Riggs \journal \PL &198B (87) 491.}

\REF\mat{S.~Matsuda \journal\PL &282B (92) 56. }

\REF\nemesha{D.~Nemeshansky \journal\PL &224B (89) 121.}

\REF\kmcon{M.~Kato and S.~Matsuda \journal\PL &172B (86) 216.}
\REF\kmnul{M.~Kato and S.~Matsuda \journal\PL &184B (87) 184.}
\REF\kmr{M.~Kato and S.~Matsuda, {\sl Advanced Studies in Pure
  Mathematics} {\bf 16} (1988),\ 205.}

\REF\schwim{A.~Schwimmer and N.~Seiberg \journal\PL &184B (87) 191.}

\REF\kmkac{M.~Kato and S.~Matsuda \journal\PTP &78 (87) 158.}

\REF\ito{K.~Ito, J.~O.~Madsen and J.~L.~Petersen \journal\PL &292B (92) 298;
{\sl Nucl. Phys.} {\bf B398} (1993), 425.}


{\centerline {\bf 1. Introduction}}
The representation theories of the $N=4$ SU(2)$_k$-extended
superconformal algebras (SCAs) have been studied extensively by now
\refmark{\yu-\mat}.
In particular,  Eguchi and Taormina\refmark{\eguchi,\taormina}
gave a comprehensive survey of
the unitary representations, whereas the superfield
formalism
was investigated systematically by Matsuda and
Uematsu\refmark{\musf,\musd}.
Also, the conjecture of the Kac determinant formulae
of the $N=4$ SU(2)$_k$ SCAs has been made by Kent and
Riggs\refmark{\kent}.
The expressions of the $N=4$ Kac formulae were obtained by a mixture of
analytic arguments and computational experiment.
Their rigorous analytic proof was left for future studies.

Recently the present author has found the Feigin-Fuchs representations
of the  $N=4$ SU(2)$_k$ SCAs expressed in terms
of the four pairs of boson and
fermion fields\refmark{\mat}.
The representations incorporate the bosonic representations of the
nonabelian SU(2)$_{\hat k}$ Kac-Moody algebras
{\it a la} Nemeshansky\refmark{\nemesha}, where we have $k=\hat k+1$.
The study of the chiral and nonchiral vertex operators
in the discovered Feigin-Fuchs representations allowed us\refmark{\mat}
to accomplish the analytic proof of
the $N=4$ Kac determinant formaulae.
In fact, in ref.8, we correctly identified the necessary
and sufficient number of the charge-screening
operators of the $N=4$ SU(2)$_k$ SCAs, thereby
proving analytically the conjectured expressions\refmark{\kent} of
the $N=4$ Kac formulae by the use of the {\it complex contour method} of
Kato and Matsuda\refmark{\kmcon,\kmnul,\kmr}.
The full analytic proof requires a rather elaborate
analysis, and the details of the study
will be presented in a future publication.

In the present paper
we report on the representation blocks of the
conformal fields,
which describe the whole representations of the $N$=4 SU(2)$_k$ SCAs.
Within the framework of the discovered
Feigin-Fuchs representations of the $N$=4 SU(2)$_k$ SCAs,
we have found a basic unit of the representation blocks which
is composed of eight \lq\lq boson-like\rq\rq
and eight \lq\lq fermion-like\rq\rq\  conformal fields
in the form of vertex operators.
By varying the
continuous parameters
which enter in the vertex operator forms of
the fundamental conformal fields in the representation blocks,
we can explore the arbitrary representations
of the $N$=4 SU(2)$_k$ SCAs including
{\it unitary} and {\it nonunitary} ones.
The transformation properties of the representation blocks
under the $N$=4 SU(2)$_k$ superconformal symmetries are
given.
Then, the whole sets of the charge-screening operators
of the $N$=4 SU(2)$_k$ SCAs
are identified out of the sixteen conformal fields
by investigating the conditions for the {\it eligible}
charge-screening operators in terms of the
continuous parameters entering in the vertex operator forms.

The immediate application of our results is to confirm that
the charge-screening operators we have obtained in ref.8 exhaust the
whole sets of them for the $N$=4 SU(2)$_k$ SCAs.
As an important consequence of this observation,
we have found\refmark{\mat} that the expressions
of the conjectured $N=4$ Kac formulae\refmark{\kent}
exhaust all the necessary factors of the determinant zeroes
to be included in the formulae.

\vskip 4cm

{\centerline{\bf 2. The $N$=4 SU(2)$_k$ superconformal algebras}}
Now, the structure of the SCAs is essentially specified by $N$ and
for each $N$ there is a range of possible modings for the algebraic
generators parametrizing distinct algebras with this basic supersymmetry
structure.\refmark{\schwim}
To get around the complexity of the moding assignments till the last
moment, we generate the $N$=4 SU(2)$_k$ SCAs
by operators $L(z), T^i(z), G^a(z),
\bar G_a(z)$ and a c-number central charge $c=6k$ in the form of operator
product expansions (OPEs):
$$ \eqalign{
  L(z)L(w)&\sim {3k\over (z-w)^4}+{2L(w)\over(z-w)^2}+
  {\partial_wL(w)\over z-w},
                                                                    \cr
&\cr
  T^i(z)T^j(w)&\sim {{1\over2}k\eta^{ij}\over(z-w)^2}+
      {{\rm i}\epsilon^{ijk}\eta_{kl}T^l(w)\over z-w},\
   \ L(z)T^i(w)\sim {T^i(w)\over (z-w)^2}+
        {\partial_wT^i(w)\over z-w},                                \cr
&\cr
  T^i(z)G^a(w)&\sim -{{1\over 2}{(\sigma^i)^a}_bG^b(w)\over z-w}\ ,\quad
    \qquad  T^i(z)\bar G_a(w)\sim
      {{1\over 2}\bar G_b(w){(\sigma^i)^b}_a\over z-w}\ ,           \cr
&\cr
  L(z)G^a(w)&\sim {{3\over2}G^a(w)\over(z-w)^2}+
     {\partial_w G^a(w)\over z-w}\ ,\
    \ L(z)\bar G_a(w)\sim{{3\over2}\bar G_a(w)\over(z-w)^2}+
         {\partial_w \bar G_a(w)\over z-w}\ ,                         \cr
&\cr
  G^a(z)G^b(w)&\sim 0\ ,\quad \bar G_a(z)\bar G_b(w)\sim 0           \cr
&\cr
  G^a(z)\bar G_b(w)&\sim {4k{\delta^a}_b\over(z-w)^3}-
     {4{(\sigma^i)^a}_b{\eta_{ij}}T^j(w)\over(z-w)^2}-
     {2{(\sigma^i)^a}_b{\eta_{ij}}\partial_w T^j(w)\over z-w}+
     {2{\delta^a}_bL(w)\over z-w}                                     \cr}
   \eqn\sca$$
The superscripts and subscripts
\ $i$=0,$\pm$\ denote SU(2) triplets in the diagonal basis,
whereas the superscripts (subscripts)\ $a$=1,2\ label
SU(2) doublet (antidoublet) representations.
The group tensors $\epsilon^{ijk},\  \eta^{ij}=\eta_{ij}$ and
the Pauli matrices ${(\sigma^i)^a}_b$
are defined in the diagonal basis.
The symmetric tensor $\eta^{ij}=\eta^{ji}$ is:
$
\ \eta^{+-}=\eta^{00}=1, etc.
$,
otherwise zero, whereas the antisymmetric tensor $\epsilon^{ijk}$ is:
$
\ \epsilon^{+-0}=-{\rm i}, etc.
$,
otherwise zero. The Pauli matrices are:
$\sigma^\pm=(\sigma^1\pm\sigma^2)/\sqrt2,\ \sigma^0=\sigma^3.$
For later convenience
the fermionic operators are hereby re-labeled
according to the canonical notation
of SU(2) spins, thus
the index $a$  runs over $\pm$
corresponding to the increase($+$) or decrease($-$) of their
third components by a half unit when the operators are applied to
conformal states.
We henceforth have
$G^-\equiv G^1,\  G^+\equiv  G^2,\  \bar G_+\equiv \bar G_1$ and
$\bar G_-\equiv \bar G_2$.
The symmetric delta function ${\delta^a}_b={\delta^b}_a$
has the standard meaning
with the SU(2) doublet or antidoublet labels, therefore  we have
$
{\delta^-}_+={\delta^1}_1=1, \ {\delta^+}_-={\delta^2}_2=1,
\ {\delta^+}_+={\delta^2}_1=0, \ {\delta^-}_-={\delta^1}_2=0
$.

\vskip .5cm

{\centerline{\bf 3. The Feigin-Fuchs representations}}
Our Feigin-Fuchs (Coulomb gas) theories of
the $N$=4 SU(2)$_k$ SCAs are represented
in terms of four real bosons
$\varphi_\alpha (z)$ ($\alpha$=1,2,3,4), and
four real fermions
forming\refmark{\musf}
a pair of complex fermion doublet $\gamma^a (z)$
and  antidoublet  $\bar\gamma_a (z)$ ($a$=1,2 or $\pm$).
The four real bosons and the pairs of complex fermions
satisfy the following OPEs :
$$\eqalign{
\varphi(z)\partial\varphi(w)\sim {1\over z-w}\ ,\qquad
\gamma^a(z)\bar\gamma_b(w)\sim {1\over z-w}{\delta^a}_b
}
   \eqn\ope$$

We first consider the currents $J^i$ ($i$=0,$\pm$)
forming the SU(2)$_{\hat k}$ Kac-Moody subalgebras
with {\it arbitrary} level $\hat k$.
Their Feigin-Fuchs
forms are given in terms of the first three bosons
by\refmark{\nemesha}
$$\eqalign{
J^0(z)     &={\rm i}{\sqrt{\hat k\over 2}}\partial\varphi_3     \cr
&\cr
J^{\pm}(z) &=\ :{{\rm i}\over \sqrt 2}
            \left( {\sqrt{\hat k+2\over 2}}\partial\varphi_1\pm
    {\rm i}{\sqrt{\hat k\over 2}}\partial\varphi_2 \right)
    e^{\pm {\rm i}{\sqrt{2\over\hat k}}(\varphi_3-{\rm i}\varphi_2)}:  \cr}
        \eqn\ff$$
The corresponding contribution of the energy-momentum tensor
is then given due to the Sugawara construction as
$${1\over \hat k+2}\sum_{i, j=0,\pm}:J^i(z)\,\eta_{ij}J^j(z):
    =-{1\over 2}\sum_{\alpha=1}^3:\left(\partial\varphi_\alpha\right)^2:+
        {\rm i}{\tau\over 2}\partial^2\varphi_1
     \eqn\sug$$
where $\tau\equiv\sqrt{2/(\hat k+2)}$.

Now, the total energy-momentum tensor $L(z)$
is given by adding the contributions from the fermion doublets and the
fourth boson to eq.{\sug} as:
$$L(z)=-{1\over 2}\sum_{\alpha=1}^4:\left(\partial\varphi_\alpha\right)^2:+
        {\rm i}{\tau\over 2}\partial^2\varphi_1-
        {\rm i}\kappa\partial^2\varphi_4+
        {1\over 2}:\left( \partial\bar\gamma\cdot\gamma-
           \bar\gamma\cdot\partial\gamma \right):
              \eqn\em$$
with the parameter value of
$\kappa\equiv{\rm i}\,(\hat k+1)\tau/2$,
where one should note the {\rm i} factor in front.

The $N$=4 supercurrents $G^a(z)$ and $\bar G_a(z)$
in our Feigin-Fuchs representations
are:
$$\eqalign{
 G^a(z) &={\rm i}\gamma^a \partial\varphi_4-
               2\kappa\partial\gamma^a-
      {\rm i}\tau J^i\,\eta_{ij}\left( \sigma^j\gamma \right)^a+
       {\rm i}\tau:\left( \bar\gamma\cdot\gamma \right)\gamma^a:    \cr
\bar G_a(z) &={\rm i}\bar\gamma_a\partial\varphi_4-
               2\kappa \partial \bar\gamma_a+
       {\rm i}\tau J^i\,\eta_{ij}\left( \bar\gamma\sigma^j \right)_a-
       {\rm i}\tau :\left( \bar\gamma\cdot\gamma \right) \bar\gamma_a:  \cr}
            \eqn\sup$$

We finally define the total SU(2)$_k$ Kac-Moody currents, $T^i(z)$, with
level $k\equiv\hat k +1$
by adding the fermionic contribution as
$$T^i(z)=J^i(z)+{1\over 2}:\bar\gamma\sigma^i \gamma(z):
   \eqn\ka$$

We note that our Feigin-Fuchs representations
just presented in fact
can be confirmed
to satisfy the $N$=4 SU(2)$_k$ SCAs, eq.{\sca},
for the parameter values being summarized as follows:
$$
\mu\equiv{\rm i}\tau={\rm i}\sqrt{2\over \hat k+2}\ ,\qquad
    \kappa={\rm i}\,(\hat k+1){\tau\over 2}\ ,\qquad
    k=\hat k+1
            \eqn\para$$
with {\it arbitrary} level $k$.

\break

{\centerline{\bf 4. Vertex operators and conformal weights}}
The vertex operators representing basic primary fields
of the $N$=4 SU(2)$_k$ SCAs are given by
$$V(t,j,j_0,z)=\ :e^{{\rm i}t\varphi_4(z)}:V_{j,\,j_0}(z)
             \eqn\vot$$
where the vertex operator $V_{j,\,j_0}(z)$ is defined as

$$V_{j,\,j_0}(z)=\ :e^{-{\rm i}j\tau \varphi_1}
   e^{{\rm i}j_0{\sqrt{2\over \hat k}}
          \left( \varphi_3-{\rm i}\varphi_2 \right)}:
\eqn\voj$$
which gives the representations labeled by SU(2)-spin
$(j,j_0)$ of
the SU(2)$_{\hat k}$ Kac-Moody algebra
and satisfy the following OPEs:
$$
J^0(z)V_{j,\,j_0}(w)\sim {j_0\over z-w}V_{j,\,j_0}(w)\ ,\qquad
\sqrt 2J^{\pm}(z)V_{j,\,j_0}(w)\sim{-j\pm j_0\over z-w}V_{j,\,j_0\pm1}(w)
      \eqn\jope$$

In our Coulomb gas representations a primary state
with conformal dimension $h$ and SU(2)-spin $(j,j_0)$ is defined as
$$ \vert h,j_0\rangle\ \sim V(t,j,j_0,z=0)\vert 0\rangle\
                           \eqn\pri$$
where $\vert 0\rangle$ is the ground state and the conformal dimension $h$
is given by
$$h=h(t,j)\equiv
{t^2\over 2}+\kappa t+{\tau^2\over 2}j(j+1)
  ={1\over 2}(t+\kappa)^2+{\tau^2\over 2}\left(j+{1\over 2}\right)^2
         +{\hat k\over 4}
     \eqn\cw$$

As discussed extensively in ref.{\kmkac}, the vertex operator
$V(-t-2\kappa,j,j_0,z)$ conjugate to $V(t,j,j_0,z)$ has the
same conformal weight
$h(-t-2\kappa,j)=h(t,j)$ as its conjugate, thus
plays crucial roles in investigating the Kac determinant formulae
in the $t$ parameter space.
The reference {\kmkac}
should be refered to for the details in this context.

Quite similarly, the vertex operator $V_{-j-1,\,j_0}(z)$
conjugate to $V_{j,\,j_0}(z)$ has the same conformal weight $\Delta_j$
as $V_{j,\,j_0}(z)$ does, with the definition of
$$\Delta_j\equiv{\tau^2\over 2}j(j+1)={j(j+1)\over \hat k+2}
   =\Delta_{-j-1}
     \eqn\jcw$$
The role of the conjugate operator has to be taken into proper account
whenever the conjugate expressions obtained by its usage
provide valid results.
Thus in our Coulomb gas representation,
for a given conformal weight $h=h(t,j)=h(t, -j-1)$,
there are, in general terms,
two ways of defining a primary state
with conformal weight $h$ and
{\it highest weight} SU(2)-spin $j_0$:
$$ \vert h=h(t,j),j_0=j\rangle\
                 \equiv  V(t,j,j_0=j,z=0)\vert 0\rangle   \eqn\pria$$
$$ \vert h=h(t, -j-1),j_0=-j-1\rangle\
                 \equiv  V(t,-j-1,j_0=-j-1,z=0)\vert 0\rangle  \eqn\prib$$

For simplicity we present the equations here for the Neveu-Schwarz case,
but our anlysis is of course general and
the results hold for other modings as well, which will be explicit later.

At this point, following ref.{\kent}, let us introduce
the ``charge'' operator $\hat C$ whose action is defined by
$$\eqalign{
\hat C\vert h,j_0\rangle & =0,\quad\left[\hat C, G_a(z)\right]=G_a(z),\quad
\left[\hat C,\bar G_a(z)\right]=-\bar G_a(z),                         \cr
&\cr
      &\quad\left[\hat C,L(z)\right]=0,\quad\left[\hat C,T^i(z)\right]=0 \quad
         \eqn\cha$$
The eigenvalues $\hat C$=$\pm 1$ actually correspond to  nothing but
the up and down of the third component of
SU(2)$_{global}$ spin.\refmark{\musf}

\vskip 5mm

{\centerline{\bf 5. Representation blocks of conformal fields}}
Next we proceed to the construction of the {\it representation blocks}
of conformal fields.
As illustrated by the rhombic dodecahedron in fig.1,
a basic unit of the representation blocks
has the top and bottom points and the
three layers in between.
The fundamental set of the conformal fields
forming  the representation blocks
can be identified
by successively taking the OPEs
of each entry in the representation blocks
with the $N=4$ SU(2)$_k$ supercurrents $G^a(z),\ \bar G_a(z) \ \ (a=\pm)$.

One starts
with the \lq\lq boson-like\rq\rq \ fundamental conformal field
sitting at the bottom point of the rhombic dodecahedron, which is
given by the basic vertex operator form $V(t,j,j_0,z)$.
By taking the OPEs of $V(t,j,j_0,z)$ with the supercurrents:
$$\eqalign{
G^{\mp}(z) V(t,j,j_0,w)&\sim {1\over z-w}V^{\mp}(t,j,j_0,w)\ ,\cr
&\cr
\bar G_{\pm}(z) V(t,j,j_0,w)&\sim {1\over z-w}\bar V_{\pm}(t,j,j_0,w)\ ,\cr}
      \eqn\opcvb$$
the four \lq\lq fermion-like\rq\rq\ {\it chiral} conformal fields,
$V^a(t,j,j_0,z)$ and $\bar V_a(t,j,j_0,z)$ $(a=\pm)$,
sitting at the four corners of the first layer
are found to be given by
$$\eqalign{
V^{\mp}(t,j,j_0,z)&=(t\mp {\rm i}\tau j_0)\gamma^{\mp}V(t,j,j_0,z)
         +{\rm i}\tau(j\pm j_0)\gamma^{\pm}V(t,j,j_0\mp 1,z)\ ,    \cr
&\cr
\bar V_{\pm}(t,j,j_0,z)&=(t\pm{\rm i}\tau j_0)\bar\gamma_{\pm}V(t,j,j_0,z)
    -{\rm i}\tau(j\mp j_0)\bar\gamma_{\mp}V(t,j,j_0\pm 1,z)\ .    \cr}
        \eqn\cv$$

The fundamental  six \lq\lq boson-like\rq\rq conformal fields,
$X^{\pm}(t,j,j_0,z),\ Y(t,j,j_0,z),\break
\bar Y(t,j,j_0,z),\ W^{(\pm)}(t,j,j_0,z)$,\
sitting singly
at the four corners of and doubly
in the center of
the middle layer of the rhombic dodecahedron are constructed by
the OPE relations:
$$\eqalign{
G^{\mp}(z) V^{\pm}(t,j,j_0,w) & \sim \pm {1\over z-w}Y(t,j,j_0,w)\ ,   \cr
&\cr
\bar G_{\pm}(z) \bar V_{\mp}(t,j,j_0,w)
                             & \sim \pm {1\over z-w}\bar Y(t,j,j_0,w)\ ,\cr}
     \eqn\y$$
$$G^{\mp}(z) V^{\mp}(t,j,j_0,w)\sim 0,\quad
  \bar G_{\pm}(z) \bar V_{\pm}(t,j,j_0,w)\sim 0
     \eqn\ynul$$
$$\eqalign{
G^{\mp}(z)\bar V_{\mp}(t,j,j_0,w) & \sim
  {2(j\pm j_0)\over (z-w)^2}V(t,j,j_0\mp 1,w)
                   -{1\over z-w}X^{\mp}(t,j,j_0,w)\ ,       \cr
&\cr
\bar G_{\pm}(z) V^{\pm}(t,j,j_0,w) & \sim
  -{2(j\mp j_0)\over (z-w)^2}V(t,j,j_0\pm 1,w)
                   +{1\over z-w}X^{\pm}(t,j,j_0,w)\ .       \cr}
    \eqn\x$$
and
$$\eqalign{
G^{\mp}(z) \bar V_{\pm}(t,j,j_0,w)  \sim
  {2(h\mp j_0)\over (z-w)^2} & V(t,j,j_0,w)                      \cr
                             &  +{1\over z-w}
   \Bigl[  \partial V(t,j,j_0,w)\pm W^{(\mp)}(t,j,j_0,w)\Bigr]\ , \cr
&\cr
\bar G_{\pm}(z) V^{\mp}(t,j,j_0,w)  \sim
   {2(h\pm j_0)\over (z-w)^2}& V(t,j,j_0,w)                       \cr
                             &  +{1\over z-w}
   \Bigl[ \partial V(t,j,j_0,w)\mp W^{(\mp)}(t,j,j_0,w) \Bigr]\ , \cr}
    \eqn\w$$
where
$$\eqalign{
Y(t,j,j_0,z) & =(t+{\rm i}\tau j)\Bigl(t-{\rm i}\tau(j+1)\Bigr)
            \gamma^-\gamma^+ V(t,j,j_0,z)\ ,                    \cr
&\cr
\bar Y(t,j,j_0,z) & =(t+{\rm i}\tau j)\Bigl(t-{\rm i}\tau(j+1)\Bigr)
     \bar\gamma_+ \bar\gamma_- V(t,j,j_0,z)\ ,                  \cr}
     \eqn\yy$$

$$\eqalign{
X^{\pm}(t,j,j_0,z)= & {\rm i}\tau(j\mp j_0)
     :\big({\rm i}\partial \varphi_4\pm {\rm i}\tau J^0\big)
                                          V(t,j,j_0\pm1,z):  \cr
  +   &{\rm i}\tau(t\pm{\rm i}\tau j_0):{\sqrt 2}J^{\pm} V(t,j,j_0,z):  \cr
  \mp & {\rm i}\tau(j\mp j_0)(t\pm {\rm i}\tau j_0)
            :{ \rm i}{\sqrt {2\over \hat k}}
         (\partial\varphi_3-{\rm i}\partial\varphi_2) V(t,j,j_0\pm 1,z): \cr
  -   &{\rm i}\tau(j\mp j_0)(t+\mu\pm {\rm i}\tau j_0)
        \big(\bar\gamma_{\mp}\gamma^{\pm}-\bar\gamma_{\pm}\gamma^{\mp}\big)
                     V(t,j,j_0\pm1,z)                                    \cr
  +   &(t\pm{\rm i}\tau j_0)(t+\mu\pm{\rm i}\tau j_0)
     \bar\gamma_{\pm}\gamma^{\pm}V(t,j,j_0,z)                            \cr
  +   &\tau^2(j\mp j_0)(j\mp j_0-1)\bar\gamma_{\mp}\gamma^{\mp}
          V(t,j,j_0\pm2,z)\ ,                                            \cr}
and
$$\eqalign{
W^{(\pm)}&(t, j,j_0,z)=\ :\Biggl[{\rm i}\tau j_0\partial\varphi_4
  -\mu t
  \Bigl({\rm i}\sqrt{\hat k\over 2}\partial\varphi_3\Bigr)
  -\tau^2j\sqrt{\hat k\over 2}\partial\varphi_2
    +{\rm i}\tau j_0\partial\varphi_1\Bigr.                           \cr
  & \hskip 2.5cm -\tau^2\Bigl(j(j+1)-j_0^2\Bigr){\rm i}\sqrt{2\over \hat k}
  (\partial\varphi_3-{\rm i}\partial\varphi_2)                        \cr
  & \pm \Bigl.(t^2+\tau^2j_0^2) \bar\gamma_{\mp}\gamma^{\pm}
  \pm\Bigl(\tau^2j(j+1)-\tau^2 j_0^2-\mu t\Bigr)\bar \gamma_{\pm}\gamma^{\mp}
             \Biggr] V(t,j,j_0,z):                                      \cr
                    \bar \gamma_+\gamma^+ V(t,j,j_0-1,z)              \cr
   &\hskip 3.5cm + {\rm i}\tau(j-j_0)(t-{\rm i}\tau j_0)
                    \bar\gamma_-\gamma^- V(t,j,j_0+1,z)\ .             \cr}
             \eqn\ww$$

The another four \lq\lq fermion-like\rq\rq\ conformal fields,
$U^a(t,j,j_0,z)$ and $\bar U_a(t,j,j_0,z)$  $(a=\pm)$,\
sitting at the four corners of the upper layer show up in the
OPE relations as
$$\eqalign{
\bar G_{\pm}(z)Y(t,j,j_0,w)&  \sim \pm {2\over (z-w)^2}
      \Bigl[  (h\pm j_0+1)  V^{\pm}(t,j,j_0,w)                        \cr
    & +(j\mp j_0) V^{\mp}(t,j,j_0\pm1,w)\Bigr]
              \pm{1\over z-w}U^{\pm}(t,j,j_0,w)\ ,                 \cr
    &\cr
               G^{\mp}(z)\bar Y(t,j,j_0,w)& \sim \pm {2\over (z-w)^2}
      \Bigl[  (h\mp j_0+1)  \bar V_{\mp}(t,j,j_0,w)                   \cr
    & -(j\pm j_0) \bar V_{\pm}(t,j,j_0\mp 1,w)\Bigr]
               \pm{1\over z-w}\bar U_{\mp}(t,j,j_0,w)\ ,                \cr }
     \eqn\ycu$$

$$\eqalign{
G^{\mp}(z)X^{\pm}(t,j,j_0,w) &  \sim
     {2\over (z-w)^2}(h\mp j_0)V^{\pm}(t,j,j_0,w)                      \cr
        & \qquad +{1\over z-w} \Bigl[2\partial V^{\pm}(t,j,j_0,w)
                                -U^{\pm}(t,j,j_0,w)\Bigr]\ ,           \cr
&\cr
\bar G_{\mp}(z)X^{\pm}(t,j,j_0,w) & \sim
    -{2\over (z-w)^2}(h\mp j_0)\bar V_{\pm}(t,j,j_0,w)                 \cr
        & \qquad -{1\over z-w} \Bigl[2\partial\bar V_{\pm}(t,j,j_0,w)
                         -\bar U_{\pm}(t,j,j_0,w)\Bigr]\ ,              \cr }
     \eqn\xcu$$
and
$$\eqalign{
G^{\mp}(z)  W^{(\pm)} &(t,j, j_0,w)\sim
   \pm{1\over (z-w)^2}\Bigl[2(j\pm j_0)V^{\pm}(t,j,j_0\mp1,w)      \cr
           & +V^{\mp}(t,j,j_0,w)\Bigr]
      \mp{1\over z-w}\Bigl[\partial V^{\mp}(t,j,j_0,w)
                   -U^{\mp}(t,j,j_0,w)\Bigr]\ ,                        \cr
&\cr
\bar G_{\pm}(z)  W^{(\pm)} &(t,j,j_0,w)\sim
   \pm{1\over (z-w)^2}\Bigl[2(j\mp j_0)\bar V_{\mp}(t,j,j_0\pm1,w)  \cr
           & -\bar V_{\pm}(t,j,j_0,w)\Bigr]
      \pm{1\over z-w}\Bigl[\partial \bar V_{\pm}(t,j,j_0,w)
                   -\bar U_{\pm}(t,j,j_0,w)\Bigr]\ ,                   \cr}
       \eqn\wcu$$
where
$$\eqalign{
U^{\mp} (t,j,j_0,z)= & (t+{\rm i}\tau j)\Bigl(t-{\rm i}\tau(j+1)\Bigr) \cr
:\biggl[ & \Bigl\{
  \gamma^{\mp}\big({\rm i}\partial\varphi_4\mp{\rm i}\tau J^0\big)
   -{\rm i}\tau\gamma^{\pm}\sqrt 2J^{\mp}
   +\mu\partial\gamma^{\mp}  \Bigr\}V(t,j,j_0,z)                       \cr
   & \quad \mp{\rm i}\tau  (j\pm j_0)\gamma^{\pm}{\rm i}{\sqrt{2\over \hat k}}
     (\partial\varphi_3-{\rm i}\partial\varphi_2)V(t,j,j_0\mp1,z)      \cr
   & \hskip 1cm +(t+\mu\mp{\rm i}\tau j_0)
           \bar\gamma_{\mp}\gamma^{\pm}\gamma^{\mp}V(t,j,j_0,z)          \cr
   & \qquad\hskip 1cm +{\rm i}\tau(j\pm j_0)
        \bar\gamma_{\pm}\gamma^{\mp}\gamma^{\pm}V(t,j,j_0\mp1,z)
              \biggr]:\ ,
                                                             \cr
&\cr
 \bar U_{\pm} (t,j,j_0,z)= & (t+{\rm i}\tau j)
                                       \Bigl(t-{\rm i}\tau(j+1)\Bigr) \cr
:\biggl[ & \Bigl\{
  \bar\gamma_{\pm}\big({\rm i}\partial\varphi_4\pm{\rm i}\tau J^0\big)
   +{\rm i}\tau\bar\gamma_{\mp}\sqrt 2J^{\pm}
   +\mu\partial\bar\gamma_{\pm}  \Bigr\}V(t,j,j_0,z)                       \cr
   & \quad \mp{\rm i}\tau  (j\mp j_0)\bar\gamma_{\mp}{\rm i}
              {\sqrt{2\over \hat k}}
     (\partial\varphi_3-{\rm i}\partial\varphi_2)V(t,j,j_0\pm1,z)      \cr
   & \hskip 1cm +(t+\mu\pm{\rm i}\tau j_0)
       \gamma^{\pm}\bar\gamma_{\mp}\bar\gamma_{\pm}V(t,j,j_0,z)          \cr
   & \qquad\hskip 1cm -{\rm i}\tau(j\mp j_0)
        \gamma^{\mp}\bar\gamma_{\pm}\bar\gamma_{\mp}V(t,j,j_0\pm1,z)
              \biggr]:\ .                                             \cr }

Let us also note the following OPE relations which are already closed
without generating the new entries,
$U^a(t,j,j_0,z)$ and $ \bar U_a(t,j,j_0,z) \ \ (a=\pm)$ :
$$G^{\mp}(z)Y(t,j,j_0,w)\sim 0\ ,\quad\qquad
\bar G_{\pm}(z)\bar Y(t,j,j_0,w)\sim 0\ ,
           \eqn\eycu$$
$$\eqalign{
G^{\pm}(z)X^{\pm}(t,j,j_0,w) & \sim
    {2(j\mp j_0)\over (z-w)^2}V^{\pm}(t,j,j_0\pm 1,w)\ ,             \cr
&\cr
\bar G_{\pm}(z)X^{\pm}(t,j,j_0,w) & \sim
    {2(j\mp j_0)\over (z-w)^2 }\bar V_{\pm}(t,j,j_0\pm 1,w)\ ,       \cr }
$$\eqalign{
G^{\pm}(z)W^{(\pm)}(t,j,j_0,w)\sim
   \pm{1\over (z-w)^2} & \Bigl[2(h\pm j_0)  +1\Bigr] V^{\pm}(t,j,j_0,w)  \cr
               & \quad\pm  {1\over z-w}\partial V^{\pm}(t,j,j_0,w)\ ,    \cr
&\cr
\bar G_{\mp}(z)W^{(\pm)}(t,j,j_0,w)\sim
   \mp{1\over (z-w)^2} & \Bigl[2(h\mp j_0) +1\Bigr]\bar V_{\mp}(t,j,j_0,w) \cr
       & \quad \mp {1\over z-w}\partial \bar V_{\mp}(t,j,j_0,w)\ .      \cr }
       \eqn\ewcu$$

Finally we proceed to the construction of the last fundamental conformal
field, $U(t,j,j_0,z)$,
sitting at the {\it top} corner of the representation block .
Taking the OPEs of the four \lq\lq fermion-like\rq\rq\  conformal fields,
$U^a(t,j,j_0,z)$ and $ \bar U_a(t,j,j_0,z)$,
in the upper layer with the supercurrents again, we generate
$U(t,j,j_0,z)$ as
$$\eqalign{
\bar G_{\pm}(z) U^{\mp}(t,j,j_0,w) & \sim
  {2\over (z-w)^2}\biggl[ (h\pm j_0+1)
         \Bigl\{ \partial V(t,j,j_0,w)\pm W^{(\pm)}(t,j,j_0,w) \Bigr\} \cr
     & \hskip 3cm +(j\mp j_0)  X^{\mp}(t,j,j_0\pm1,w) \biggr]        \cr
  +{1\over z-w} &
  \biggl[ \partial W^{(+)}(t,j,j_0,w)-\partial W^{(-)}(t,j,j_0,w)
               + U(t,j,j_0,w)\biggr]\ ,                                 \cr
& \cr
G^{\mp}(z)\bar U_{\pm}(t,j,j_0,w) & \sim
  {2\over (z-w)^2}\biggl[ (h\mp j_0+1)
        \Bigl\{\partial V(t,j,j_0,w)\mp W^{(\pm)}(t,j,j_0,w) \Bigr\} \cr
      & \hskip 3cm +(j\pm j_0)  X^{\pm}(t,j,j_0\mp1,w) \biggr]      \cr
  -{1\over z-w} &
  \biggl[ \partial W^{(+)}(t,j,j_0,w)-\partial W^{(-)}(t,j,j_0,w)
               - U(t,j,j_0,w)\biggr]\ .                                 \cr}
      \eqn\u$$
where the {\it celebrated} final entry
$U(t,j,j_0,z)$
in the representation block is given by \nobreak
the extremely elaborate expression as follows:         \hfil

$$\eqalign{
 U & (t,j,j_0,z)= (t+{\rm i}\tau j)\Bigl(t-{\rm i}\tau(j+1)\Bigr)     \cr
& :\Biggl[ (t-{\rm i}\tau j)
      \Bigl(t+{\rm i}\tau(j+1)\Bigr)
  \bar \gamma_+\gamma^-\bar \gamma_-\gamma^+V(t,j,j_0,z)
                                                                      \cr
&\qquad +(1-\mu t)
    \Bigl(\partial\bar\gamma_+\cdot\gamma^-
         +\partial\bar\gamma_-\cdot\gamma^+
    - \bar\gamma_+\cdot\partial\gamma^-
              -\bar\gamma_-\cdot\partial\gamma^+\Bigr)
                                               V(t,j,j_0,z)            \cr
 & +\tau^2(j+j_0)
    \partial\big(\bar\gamma_+\cdot\gamma^+\big)V(t,j,j_0-1,z)
  +\tau^2(j-j_0)\partial\big(\bar\gamma_-\cdot\gamma^-\big)V(t,j,j_0+1,z)
                                                                     \cr
 &\hskip 2cm +{\rm i}\tau\mu j_0\partial
    \big(\bar\gamma_+\cdot\gamma^--\bar\gamma_-\cdot\gamma^+\big)V(t,j,j_0,z)
    \Biggr.                                                     \cr
 &\qquad +\Bigl(-\sum_{\alpha=1}^4\big(\partial\varphi_\alpha\big)^2
      +{\rm i}\tau\partial^2\varphi_1
         +{\rm i}\mu\partial^2\varphi_4
         \Bigr)V(t,j,j_0,z)                               \cr
 &\hskip 5cm    +\tau^2j_0{\rm i}{\sqrt{2\over \hat k}}
          \big(\partial^2\varphi_3
        -{\rm i}\partial^2\varphi_2\big)V(t,j,j_0,z)
                                                                       \cr
 &\qquad  +{\rm i}{\sqrt{2\over \hat k}}
           \big(\partial\varphi_3 -{\rm i}\partial\varphi_2\big)
        \Bigl( \tau^2(j +j_0){\sqrt 2}J^+V(t,j,j_0-1,z)             \cr
 &\hskip 5cm   -\tau^2(j-j_0)  \sqrt 2J^-V(t,j,j_0+1,z)\Bigr)          \cr
 &\qquad  -\Bigl(\tau^2j(j+1)-\tau^2j_0^2\Bigr)
         \Bigl({\rm i}{\sqrt{2\over \hat k}}\big(\partial\varphi_3
         -{\rm i}\partial\varphi_2\big)\Bigr)^2V(t,j,j_0,z)          \cr
 &\quad   +\big(\bar\gamma_+\gamma^--\bar\gamma_-\gamma^+\big)
       \Bigl\{
   -{1\over 2}(t-{\rm i}\tau j_0)
         \big({\rm i}\partial\varphi_4+{\rm i}\tau J^0\big)V(t,j,j_0,z)
                                                                   \cr
 & \hskip 4.5 cm  +{1\over 2}(t+{\rm i}\tau j_0)
        \big({\rm i}\partial\varphi_4-{\rm i}\tau J^0\big)V(t,j,j_0,z)
                                                                     \cr
 &\quad   -{1\over 2}\tau^2(j+j_0)
           {\sqrt 2}J^+  V(t,j,j_0-1,z)
    +{1\over 2}\tau^2(j-j_0){\sqrt 2}J^-V(t,j,j_0+1,z)
                                                                 \cr
 &\hskip 2 cm   +\Bigl(\tau^2j(j+1)-  \tau^2j_0^2)\Bigr)
       {\rm i}{\sqrt{2\over \hat k}}
      \big(\partial\varphi_3-{\rm i}\partial\varphi_2\big)V(t,j,j_0,z)
           \Bigr\}
                                                                  \cr
&   -{\rm i}\tau(t-{\rm i}\tau j_0)
         {\sqrt 2}J^+
       \bar\gamma_-\gamma^-V(t,j,j_0,z)
    -{\rm i}\tau(t+{\rm i}\tau j_0){\sqrt 2}J^-
       \bar\gamma_+\gamma^+V(t,j,j_0,z)                                \cr
&\hskip 2cm  \Biggl.   -{\rm i} \tau(j-j_0)
        \big({\rm i}\partial\varphi_4
            -{\rm i}\tau J^0\big)
        \bar\gamma_-\gamma^-V(t,j,j_0+1,z)                       \cr
&\hskip 2cm    -{\rm i}  \tau(j+j_0)
         \big({\rm i}\partial\varphi_4
              +{\rm i}\tau J^0\big)
        \bar\gamma_+\gamma^+V(t,j,j_0-1,z)                           \cr
&\quad\qquad    +{\rm i}\tau(j-j_0)(t-{\rm i}\tau j_0)
        {\rm i}{\sqrt{2\over \hat k}}
        \big(\partial\varphi_3-{\rm i}\partial\varphi_2\big)
               \bar\gamma_-\gamma^-V(t,j,j_0+1,z)                     \cr
&\quad\qquad     -{\rm i}\tau(j+j_0)(t+{\rm i}\tau j_0)
        {\rm i}{\sqrt{2\over \hat k}}
         \big(\partial\varphi_3-{\rm i}\partial\varphi_2\big)
           \bar\gamma_+\gamma^+V(t,j,j_0-1,z)     \Biggr]:\ .          \cr
              }
      \eqn\udef$$

The other combinations of the
OPE operations of the supercurrents on\break
$U^a(t,j,j_0,z)$ and
$ \bar U_a(t,j,j_0,z)$
do not produce $U(t,j,j_0,z)$, but are given by
$$\eqalign{
\bar G_{\mp}(z)  U^{\mp} & (t,j,j_0,w)\sim
       -{2\over (z-w)^2}\biggl[
      (h \mp j_0+1)X^{\mp}(t,j,j_0,w)                           \cr
   &   +(j\pm j_0)\Bigl\{\partial V(t,j,j_0\mp 1,w)
          \pm W^{(\pm)}(t,j,j_0\mp 1,w)\Bigr\} \biggr]            \cr
  &\cr
 G^{\mp}(z) \bar U_{\mp} & (t,j,j_0,w)\sim
       {2\over (z-w)^2}\biggl[
      (h \mp j_0+1)X^{\mp}(t,j,j_0,w)                           \cr
   &   +(j\pm j_0)\Bigl\{\partial V(t,j,j_0\mp 1,w)
         \pm W^{(\mp)}(t,j,j_0\mp 1,w)\Bigr\} \biggr]            \cr
               }
       \eqn\guw$$
$$\eqalign{
G^{\mp}(z)U^{\mp}(t,j,j_0,w)  & \sim
   \mp{2\over(z-w)^2}(j\pm j_0)Y(t,j,j_0\mp 1,w)                \cr
&\cr
\bar G_{\mp}(z)\bar U_{\mp}(t,j,j_0,w)  & \sim
   \pm{2\over(z-w)^2}(j\pm j_0)\bar Y(t,j,j_0\mp 1,w)            \cr
   }
   \eqn\guy$$
$$\eqalign{
G^{\pm}(z)U^{\mp}(t,j,j_0,w)\sim
  \mp{2\over (z-w)^2}(h\pm j_0)Y(t,j,j_0,w)
  \mp\partial_w\biggl({2\over z-w}Y(t,j,j_0,w)\biggr)   \cr
&\cr
\bar G_{\mp}(z)\bar U^{\mp}(t,j,j_0,w)\sim
  \pm{2\over (z-w)^2}(h\pm j_0)\bar Y(t,j,j_0,w)
  \pm\partial_w\biggl({2\over z-w}\bar Y(t,j,j_0,w)\biggr)   \cr
 }
  \eqn\gupy$$

Let us also note here that
further operations of the supercurrents $G^a(z),\ \bar G_a(z)\break
(a=\pm)$\
on $U(t,j,j_0,z)$ do not
generate any new entry, but only produce the following closed OPE
relations:
$$\eqalign{
G^{\mp}(z) & U(t,j,j_0,w)\sim
 {2\over (z-w)^3}\biggl[(h\mp j_0+1)V^{\mp}(t,j,j_0,w)      \cr
&  +(j\pm j_0) V^{\pm}(t,j,j_0\mp1,w)\biggr]
  +{2\over (z-w)^2}\biggl[\Bigl(h\mp j_0+{3\over 2}\Bigr)
               U^{\mp}(t,j,j_0,z)                        \cr
&\qquad    +(j\pm j_0) U^{\pm}(t,j,j_0\mp1,w)\biggr]
             + {1\over z-w}\partial U^{\mp}(t,j,j_0,w)\ ,    \cr
&
   \cr
\bar G_{\pm}(z) & U(t,j,j_0,w)\sim
 {2\over (z-w)^3}\biggl[(h\pm j_0+1)\bar V_{\pm}(t,j,j_0,w)    \cr
 & -(j\mp j_0)\bar V_{\mp}(t,j,j_0\pm1,w)\biggr]
  +{2\over (z-w)^2}\biggl[\Bigl(h\pm j_0+{3\over 2}\Bigr)
                       \bar U_{\pm}(t,j,j_0,z)               \cr
& \qquad   -(j\mp j_0)\bar U_{\mp}(t,j,j_0\pm1,w)\biggr]
               + {1\over z-w}\partial\bar U_{\pm}(t,j,j_0,w)\ .    \cr
      }
      \eqn\gu$$

We have thus completed the assignment of the fundamental sets
of the eight \lq\lq fermion-like\rq\rq and eight
\lq\lq boson-like\rq\rq conformal fields
to the fourteen {\it nondegenerate} corners
and the one {\it doubly degenerate} center
of the rhombic dodecahedron forming
the representation blocks.

Now let us mention the transformation properties under $L(z)$
of each entry
of the fundamental conformal fields in the representation blocks.
The  three \lq\lq boson-like\rq\rq\  fields,
$V(t,j,j_0,z),\ Y(t,j,j_0,z)$ and $\bar Y(t,j,j,_0,z),$
and the four
\lq\lq fermion-like\rq\rq\  fields,
$V^a(t,j,j_0,z)$ and $\bar V_a(t,j,j_0,z)\ \ (a=\pm),$
transform as {\it primary} fileds, i.e.
$$\eqalign{
L(z)V(t,j,j_0,w) & \sim
   {1\over z-w}\partial V(t,j,j_0,w)
   +{1\over (z-w)^2}hV(t,j,j_0,w)                                   \cr
& \cr
L(z)V^a(t,j,j_0,w) & \sim
   {1\over z-w}\partial V^a(t,j,j_0,z)
   +{1\over (z-w)^2} \Bigl(h+{1\over 2}\Bigr)V^a(t,j,j_0,w)         \cr
& \cr
L(z)\bar V_a(t,j,j_0,w) & \sim
   {1\over z-w}\partial \bar V_a(t,j,j_0,z)
   +{1\over (z-w)^2} \Bigl(h+{1\over 2}\Bigr)\bar V_a(t,j,j_0,w)    \cr
& \cr
L(z)Y(t,j,j_0,w) & \sim
  {1\over z-w}\partial Y(t,j,j_0,w)+ {1\over (z-w)^2}(h+1)Y(t,j,j_0,w) \cr
& \cr
L(z)\bar Y(t,j,j_0,w) & \sim
  {1\over z-w}\partial \bar Y(t,j,j_0,w)
  + {1\over (z-w)^2}(h+1)\bar Y(t,j,j_0,w)\ .                        \cr
}
      \eqn\lpr$$
The conformal fields $W^{(\pm)}(t,j,j_0,z)$ satsify the OPEs as
$$\eqalign{
L(z)W^{(\pm)}(t,j,j_0,w) & \sim
{1\over z-w}\partial W^{(\pm)}(t,j,j_0,w)                          \cr
 +{1\over (z-w)^2} & (h+1)W^{(\pm)}(t,j,j_0,w)
-{2\over (z-w)^3}j_0 V(t,j,j_0,w)\ .                                \cr
}
        \eqn\lw$$
Thus we find that the difference of the two fields,
$$\eqalign{
W^{(+)}(t,j,j_0,z) & -W^{(-)}(t,j,j_0,z)=                         \cr
                   & (t+{\rm i}\tau j)\Bigl(t-{\rm i}\tau(j+1)\Bigr)
\big(\bar\gamma_+\gamma^-+\bar\gamma_-\gamma^+\big)
            V(t,j,j_0,w)\ ,                                       \cr
 }
     \eqn\wdif$$
being the fourth \lq\lq boson-like\rq\rq \ {\it primary}
combination,
satisfy the primary field conditon with conformal weight $h+1$.

The other \lq\lq boson-like\rq\rq\ conformal fields,
$X^{\pm}(t,j,j_0,z)$, satsify
$$\eqalign{
L(z)   X^{\pm}(t,j,j_0,w)\sim
{1\over z-w} & \partial X^{\pm}(t,j,j_0,w)
    +{1\over (z-w)^2}(h+1)X^{\pm}(t,j,j_0,w)                    \cr
        &   -{2\over (z-w)^3}(j\mp j_0)V(t,j,j_0\pm1,w)        \cr
  }
     \eqn\lx$$

The OPE operation of $L(z)$ on the four
\lq\lq fermion-like\rq\rq   fields,
$U^a(t,j,j_0,z)$ and $\bar U_a(t,j,j_0,z)$,
being assigned to the upper layer
gives
$$\eqalign{
L(z) & U^{\mp}(t,j,j_0,w)\sim
   {1\over z-w}\partial U^{\mp}(t,j,j_0,w)
       +{1\over (z-w)^2}\Bigl(h+{3\over 2}\Bigr)U^{\mp}(t,j,j_0,w)   \cr
   & +{2\over (z-w)^3}\biggl[(h\mp j_0+1)V^{\mp}(t,j,j_0,w)
   +(j\pm j_0)V^{\pm}(t,j,j_0\mp1,w)\biggr]                      \cr
&\cr
L(z) & \bar U_{\pm}(t,j,j_0,w)\sim
   {1\over z-w}\partial \bar U_{\pm}(t,j,j_0,w)
       +{1\over (z-w)^2}\Bigl(h+{3\over 2}\Bigr)\bar U^{\pm}(t,j,j_0,w)   \cr
   & +{2\over (z-w)^3}\biggl[(h\pm j_0+1)\bar V^{\pm}(t,j,j_0,w)
   +(j\mp j_0)\bar V^{\mp}(t,j,j_0\pm1,w)\biggr]                      \cr
   }
   \eqn\lcu$$

Finally, the OPE relation of $L(z)$ with the top entry,
$U(t,j,j_0,z),$ in the rhombic dodecahedron of the representation block
is given by
$$\eqalign{
L(z) & U(t,j,j_0,w)\sim
   {1\over z-w}\partial U(t,j,j_0,w)+{1\over (z-w^2}(h+2)U(t,j,j_0,w)   \cr
   & +{2\over (z-w)^3}\Bigl[(j+j_0)\bar\gamma_+\gamma^+V(t,j,j_0-1,w)
       +(j-j_0)\bar\gamma_-\gamma^-V(t,j,j_0+1,w)                       \cr
   &\hskip 3cm  -j_0\big(\bar\gamma_+\gamma^--\bar\gamma_-\gamma^+\big)
         V(t,j,j_0,w)\Bigr]                                          \cr
         }
         \eqn\lu$$

{\centerline {\bf 6. Identification of Charge-Screening Operators}
The results in the previous section allow us to identify the
whole sets of the charge-screening operators of the $N=4$ SU(2)$_k$ SCAs.
In the following we shall explain how the
whole sets of the  screening operators
with conformal dimension one can be constructed.

The basic requirements for an {\it eligible} charge-screening operator
${\cal S}(z)$
are that it transforms as a primary field with conformal dimension one
and that it stays invariant under the transformaions of
the $N=4$ SU(2)$_k$ SCAs.
Operationally these requirements are satisfied
if and only if the following OPE relations
$$ {\cal O}(z){\cal S}(w)\sim
   \partial_w\biggl[{1\over z-w}{\cal Q}(w)\biggr]
   \quad   {\rm or} \quad
            {\cal O}(z){\cal S}(w) \sim 0
   \eqn\scr$$
should be valid for  any generator ${\cal O}(z)$ of the $N=4$ SU(2)$_k$ SCAs
with $ {\cal Q}(z)$ being a certain nonzero operator or just being zero.
In particular,  the following primary field
OPE relation should hold
$$L(z){\cal S}(w)\sim \partial_w\biggl[{1\over z-w}{\cal S}(w)\biggr]
   \eqn\pri$$
for any {\it eligible} charge-screening operator ${\cal S}(z)$.

The OPE relations we have obtained in the previous section allow us
to study unambiguously
whether any of the sixteen conformal fields assigned to the
rhombic dodecahedron could be an {\it eligible} candidate for
the charge-screening operators of the $N=4$ SU(2)$_k$ SCAs.

\noindent{\bf 6.1. Nonchiral screening operator}

As an illustration we shall start with giving the
{\it nonchiral}  charge-screening  operator ${\cal N}(z)$
which can be constructed out of the {\it celebrated} top entry
$U(t,j,j_0,z)$ in the representation blocks.
In order for $U(t,j,j_0,z)$ to be a screening operator
it has to be primary.
Therefore, from eq.{\lu} we obtain the parameter conditions
for $U(t,j,j_0,z)$:
\ $h(t,j)+2=1,\ j+j_0=0,\ j-j_0,\ j_0=0$.
Thus $j=j_0=0$ and $h=h(t,j=0)=-1$,
whose last equation determines $t$:\ $t=\mu$ or $t=-2\kappa-\mu$.
The former solution is excluded because $U(t=\mu,j=0,j_0=0,z)$
trivially vanishes, while the second solution could correspond to
a possible candidate for a charge-screening operator.
In fact, $U(t=-2\kappa-\mu, j=0,j_0=0,z)$ is an {\it eligible}
charge-screening operator, which we call the
{\it nonchiral screening operator} ${\cal N}(z)$ \refmark{\mat}
and is given by
$$\eqalign{
   {\cal N}(z)& \equiv  U(t=-2\kappa-\mu, j=0,j_0=0,z)                  \cr
              & =  2(\kappa+\mu)(2\kappa+\mu)
    \ :\Bigl[ \left({\rm i}\partial\varphi_4 \right)^2
+{\rm i}\mu\partial^2\varphi_4
+\kappa(2\kappa+\mu)\left(\bar\gamma\cdot\gamma\right)^2    \Bigr.  \cr
   -(  \partial & \bar\gamma
      \cdot\gamma -\bar\gamma\cdot\partial\gamma)
                  +\tau^2J^i\eta_{ij}J^j
       -2(\bar\gamma\sigma^i\gamma)\eta_{ij} J^j \Bigr]
        V(t=-2\kappa-\mu,j=0,j_0=0,z):                              \cr}
Its OPE's with the $N$=4 generators  either vanish  or turn into
total derivatives.\refmark{\kmr}  To save space, we only present here
the nontrivial relations which we obtain from
eqs.{\gu} and {\lu}:
$$\eqalign{
L(z){\cal N}(w)& \sim \partial_w\left[{1\over z-w}{\cal N}(w)\right]    \cr
G^a(z){\cal N}(w)&\sim \partial_w\left[{1\over z-w}
  U^a(t=-2\kappa-\mu,j=0,j_0=0,w)
         \right]                                        \cr}
                              \eqn\opnscr$$
The similar OPE relation holds for $\bar G_a(z){\cal N}(w)$ as well.

The  nonchiral  screening  operator ${\cal N}(z)$
basically generates those null states with the same
isospin and ``charge "quantum numbers
as the primary highest weight state. Only the mass level $n$ is raised
relative to the primary ground state.

\noindent{\bf 6.2. SU(2)$_\bfk$ Kac-Moody screening operator}

Next we shall consider the possibility of getting an {\it eligible}
charge-screening operator from $X^{\pm}(t,j,j_0,z)$.
In order for $X^{\pm}(t,j,j_0,z)$ to be primary,
we have to have $h(t,j)+1=1,\ j\mp j_0=0$ from eq.{\lx}.
Also, in eqs.{\xcu} and {\excu} we have to satisfy the relations
$h(t,j)\mp j_0=1,\ j\mp j_0=0$, and
$(t+{\rm i}\tau j)\big(t-{\rm i}\tau(j+1)\big)=0$,
so that $U^a(t,j,j_0,z)\equiv 0,\ \bar U_a(t,j,j_0,z)\equiv 0$ and
consequently the OPE relations $G^a(z)X^{\pm}(t,j,j_0,w)$ and
$\bar G_a(z)X^{\pm}(t,j,j_0,w)$ either
turn into total derivatives or vanish.
The nontrivial solution exists only for the choice
$\big(t-{\rm i}\tau(j+1)\big)=0$ with the solution
$t=0, j=-1, j_0=\mp 1$.
The other choice $(t+{\rm i}\tau j)=0$ gives the
trivially vanishing solution $X^{\pm}(t=0,j=0,j_0=0,z)\equiv 0$
as a result of the constraints $ t=0, j=j_0=0$,
henceforth is excluded.

We thus find
another ``charge" preserving screening operators to be given
by\refmark{\mat}
$$\eqalign{
{\cal J}_{(\pm)}(z)\equiv & \mp X^{\pm}(t=0, j=-1, j_0=\mp 1, z)
                       = \tau^2  :J^{\pm}(z)V_{j=-1,\,j_0=\mp 1}(z):     \cr
            =&\ \tau^2 :{\rm i}\left({1\over \tau}\partial\varphi_1
     \pm{\rm i}\sqrt{\hat k\over 2}\partial\varphi_2\right)
       e^{{\rm i}\tau\varphi_1}:                                 \cr}
          \eqn\kscr$$
whose OPE relations  with the $N$=4 generators  either are
trivially zero or become nontrivial total derivatives.
Only some of the nontrivial OPEs  are presented here:
$$\eqalign{
&J^{\mp}(z){\cal J}_{(\pm)}(w)\sim \partial_w\left[
  {1\over z-w}V_{-1,\mp1}(w)\right]\ ,     \cr
&G^a(z){\cal J}_{(\pm)}(w)\sim \partial_w\left[
  {1\over z-w}(\sigma^{\pm}\gamma)^a(w)V_{-1,\mp1}(w)\right]   \cr}
      \eqn\opkscr$$
with the similar OPE relations for $\bar G_a(z){\cal J}_{(\pm)}(w)$
being valid.
We remark here that we have in effect a single SU(2)$_k$ Kac-Moody
screening operator ${\cal J}(z)$:
$${\cal J}(z)\equiv {\cal J}_{(+)}(z)
              \simeq -{\cal J}_{(-)}(z)
     \eqn\j$$
since the sum ${\cal J}_{(+)}(z)+{\cal J}_{(-)}(z)$
is just a total derivative.

By the use of the similar consideration for $W^{(\pm)}(t,j,j_0,z)$
we also obtain
another {\it eligible} charge-screening operator ${\cal W}(z)$,
which is actually equivalent to ${\cal J}_{(\pm)}(z)$ as follows:
$$\eqalign{
{\cal W}(z)  \equiv & W^{(\pm)}(t=0,j=-1,j_0=0,z)
     =\ \tau^2:\sqrt{\hat k\over 2}\partial \varphi_2
         e^{{\rm i}\tau\varphi_1}:                              \cr
                   =& \pm{\cal J}_{(\mp)}(z)
         \mp\partial V(t=0,j=-1,j_0=0,z)
         \simeq -{\cal J}(z)       \cr}
         \eqn\scw$$
As for ${\cal W}(z)$ we have
the following OPEs in the form of total derivatives:
$$\eqalign{
G^a(z){\cal W}(z)
  & \sim \mp\partial_w  \biggl[{1\over z-w}V^a(t=0,j=-1,j_0=0,w)
                                                           \biggr],\cr
&  \cr
\bar G_a(z){\cal W}(z)
  & \sim\pm\partial_w\biggl[{1\over z-w}
          \bar V_a(t=0,j=-1,j_0=0,w)\biggr]                      \cr}
          \eqn\wt$$

We remark that the conformal fields $Y(t,j,j_0,z)$ and
$\bar Y(t,j,j_0,z)$ do not give rise to any eligible charge-screening
operators, which can be confirmed from eqs. {\lpr}, {\ycu} and {\eycu}.
The only allowed solution
$t=0, j=-1, j_0=0$ for the parameters
being constrained by the eligible conditions
for charge-screening operators
gives the
trivially vanishing results such as $Y(t=0,j=-1,j_0=0,z)\equiv 0$
and $\bar Y(t=0,j=-1,j_0=0,z)\equiv 0$.
Similarly, the conformal field $V(t,j,j_0,z)$
cannot satisfy the conditions
eq.{\scr} for an {\it eligible} screening operator
except in the form of the trivial
solution $V(t=0,j=0,j_0=0,z)=1$.

\noindent{\bf 6.3. Chiral screening operators}

Next we shall proceed to constructing   \lq\lq  fermion-like\rq\rq\
charge-screening operators.
The eligible candidates for them can be
obtained by studying
the \lq\lq  fermion-like\rq\rq\
conformal fields $V^a(t,j,j_0,z),$ $ \bar V_a(t,j,j_0,z),$
$U^a(t,j,j_0,z)$ and $\bar U_a(t,j,j_0,z)$.

First we consider the conformal fields
$V^{\mp}(t,j,j_0,z)$.
In this case there is some judicious investigation involved, since
the primary condition  eq.{\lpr} for
an {\it eligible} charge-screening operator
just gives the constraint on the conformal dimension such as
$h(t,j,)=1/2$ and nothing else.
{}From eqs.{\y}$\sim${\ww}
we get additional  constraints for the parameters
$t, j, j_0$ such that the equalities
$Y(t,j,j_0,z)=0, j\pm j_0=1, X^{\mp}(t,j,j_0,z)=-2\partial V(t,j,j_0\mp1,z),
h(t,j)\pm j_0=1, W^{(\mp)}(t,j,j_0,z)=\mp \partial V(t,j,j_0,z)$
should hold.
The first equality provides the constraint
$(t+{\rm i}\tau j)\big(t-{\rm i}\tau(j+1)\big)=0$.
Remarkably enough, these constraints have  nontrivial solutions:
\ $j=1/2, j_0=\pm 1/2, t=-{\rm i}\tau j=-{\rm i}\tau/2,
h(t=-{\rm i}\tau j,j)=j=1/2$ due to the following identities
$$\eqalign{
 &X^{\mp}\Bigl(t=-{\rm i}{\tau\over 2},
             j={1\over 2},j_0=\pm{1\over 2},z\Bigr)
  =-2\partial V\Bigl(t=-{\rm i}{\tau\over 2},
              j={1\over 2},j_0=\mp{1\over 2},z\Bigr)\ ,
                                         \cr
 &\cr
 &W^{(\mp)}\Bigl( t=-{\rm i}{\tau\over 2},
             j={1\over 2},j_0=\pm{1\over 2},z\Bigr)
   =\mp\partial V\Bigl(t=-{\rm i}{\tau\over 2},
               j={1\over 2},j_0=\pm{1\over 2},z\Bigr)\ .
                                       \cr}
         \eqn\xv$$
We thus find
the {\it eligible  chiral} screening operators
${\cal V}^{(\mp)}(z)$  carrying  the
``charge" of $\hat C$=$1$
to be given in terms of the chiral vertex operators, eq.{\cv},
as follows:
$${\cal V}^{(\mp)}(z) \equiv
V^{\mp}\Bigl(t=-{\rm i}{\tau\over 2},j={1\over 2},j_0=\pm{1\over 2},z\Bigr)
    =-{\cal V}^{(\pm)}(z)\equiv \mp {\cal C}(z) ,
           \eqn\cs$$

By making a similar investigation for
$\bar V_{\pm}(t,j,j_0,z)$,
we get
the {\it anti-chiral} screening operators
with the ``charge" of $\hat C$=$-1$
as
$$\bar{\cal V}_{(\pm)}(z) \equiv
\bar V_{\pm}\Bigl(t=-{\rm i}{\tau\over 2},j={1\over 2},
                 j_0=\mp{1\over 2},z\Bigr)
    =\bar{\cal V}_{(\mp)}(z)\equiv \bar{\cal C}(z) ,
           \eqn\acs$$

One should note that the ``charge" conjugate expressions of the
above chiral vertex operators
identically vanish:
$$  \bar V_{\pm}\Bigl(t=-{\rm i}{\tau\over 2},
               j={1\over 2},j_0=\pm{1\over 2},z\Bigr)
    \equiv 0 ,\qquad
   V^{\mp}\Bigl(t=-{\rm i}{\tau\over 2},
              j={1\over 2},j_0=\mp{1\over 2},z\Bigr)
    \equiv 0 .
                   \eqn\vcscr$$

Here again the
OPEs of ${\cal C}(z)$ and $\bar{\cal C}(z)$
with the $N$=4 generators either vanish or
are expressed as total derivatives.
In particular, ${\cal C}(w)$ is annihilated by $G^a(z)$ while
$\bar{\cal C}(w)$ by $\bar G_a(z)$,
that is,
$$G^a(z){\cal C}(w)\sim 0, \qquad \bar G_a(z)\bar{\cal C}(w)\sim 0.
   \eqn\gcnul$$
We present here some of the nontrivial relations taking the form of
total derivatives:
$$\bar G_{\pm}(z){\cal C}(w)  \sim \mp 2\partial_w
    \left[{1\over z-w}V\Bigl(t=-{\rm i}{\tau\over 2},j={1\over 2},
    j_0=\pm{1\over 2},w\Bigr)\right] ,
        \eqn\opcscr$$
whose ``charge" conjugate OPE relations
for $G^{\mp}(z)\bar{\cal C}(w)$
also hold in a similar form.

We finally remark that the \lq\lq fermion-like\rq\rq\  conformal fields
$U^a(t,j,j_0,z)$ and $\bar U_a(t,j,j_0,z)$
do not give rise to any eligible charge-screening operators.
The only allowed solutions $t=-{\rm i}\tau j={\rm i}\tau/2,
j=-1/2, j_0=\pm 1/2, h(t=-{\rm i}\tau j,j)=j=-1/2$ for
$U^{\mp}(t,j,j_0,z)$ to be  eligible screening operators
satisfy
the constraint $(t+{\rm i}\tau j)\big(t-{\rm i}\tau(j+1)\big)=0$,
which in turn kills in a trivial manner
not only $U(t,j,j_0,z)\equiv 0$ and
$Y(t,j,j_0,z)\equiv 0$, but also $U^{\mp}(t,j,j_0,z)\equiv 0$ itself.
Similarly this is the case for $\bar U_a(t,j,j_0,z)$.

To summarize, those  {\it eligible} charge-screening operators that
we have obatained in the above are conformally invariant
vertex operators.

\vskip .5cm

\centerline{{\bf 7. Comments and conclusions}}

We have some comments in order.
In the two illustrations 2-(a) and 2-(b)
of fig.2  are shown the steps of the OPE operations
for the fundamental conformal field
$V(t,j,j_0,z)$ sitting at the bottom of the rhombic dodecahedron
to finally reach the top point, thus generating the {\it celebrated}
$U(t,j,j_0,z)$ field.
First of all, the corresponding OPE operations
in figs.2-(a) and 2-(b) are
orthogonal to each other.
Therefore, the two illustrations are at right angles
to each other as a whole.
Also, the OPE operations in the upper rectangle in one illustration
are twisted at
right angles from those in the lower rectangle in the same illustration.
That is,
the OPE operations on the conformal fields $W^{(\pm)}(t,j,j_0,z)$
to let them climb one floor up from the middle layer
generating the \lq\lq fermion-like\rq\rq\  fields,
$U^a(t,j,j_0,z)$ and $\bar U_a(t,j,j_0,z)$, are orthogonal
to those OPE operations which
have generated $W^{(\pm)}(t,j,j_0,z)$ themselves
from $V(t,j,j_0,z)$ sitting at the bottom point.

In spite of the complicated structure of the sixteen conformal fields
assigned to the fourteen edge corners and
the one degenerate center  of the rhombic dodecahedron,
they possess a neat substructure representing the underlying
subalgebras of the $N$=4 SU(2)$_k$ SCAs,
which is the $N$=2 U(1) SCAs.

We first note that the two subsets of the operators,
$\{L(z), G^+(z), \bar G_-(z), T^0(z)\}$ and
$\{L(z), \bar G^+(z), G_-(z), T^0(z)\}$,
in the $N$=4 SU(2)$_k$ SCAs
respectively form the $N$=2 U(1) SCAs.
Consequently, the two correspondingly chosen
subsets of the \lq\lq boson-like\rq\rq\
and \lq\lq fermion-like\rq\rq \ conformal fields,
$\{V(t,j,j_0,z), V^+(t,j,j_0,z),
\break
\bar V_-(t,j,j_0,z),
W^{(+)}(t,j,j_0,z)\}$ and
$\{V(t,j,j_0,z), \bar V_+(t,j,j_0,z),
V^-(t,j,j_0,z),
\break
W^{(-)}(t,j,j_0,z)\},$
respectively stand\refmark{\kmnul,\kmr}
for the complete representations of
the $N$=2 U(1) SCAs
which are given by the above mentioned
two subsets of the generators.
In the generic notation of ref.{\kmr}
each set of these conformal fields
forms a {\it closed} primary multiplet $\{F(z), G^{\pm}(z), H(z)\}$
representing the $N$=2 U(1) SCAs.
The closed property can be confirmed from eq.\ewcu,
where no new entries are generated by the OPE operations,
that is, the OPE relations are closed among the primary multiplets
under the corresponding $N$=2 U(1) SCAs.
In fig.2 this is shown by the orthogonality of the OPE operations
in the lower rectangles to those in the upper ones.
As already stated in the above,
the supercharge generators of the $N=2$ U(1) SCAs in the lower rectangles
of fig.2(a) and fig.2(b) respectively
do not generate any new conformal field
in the directions parallel to the lower rectangles,
when they are applied to the conformal fields $W^{(\pm)}(t,j,j_0,z)$.
One set of the supercharge generators forming
the  $N=2$ U(1) subalgebra has to switch over to another set
of the supercharge generators in the orthogonal rectangle
in order to generate new entries by those OPE operations
on $W^{(\pm)}(t,j,j_0,z)$.

Now, in conclusion,
we have the necessary and sufficient tools of accomplishing the
complete study of the representation theory of
the $N$=4 SU(2)$_k$ SCAs.
For example, the fundamental unitary representations of
the $N$=4 SU(2)$_k$ SCAs constructed in ref.{\eguchi}
can be studied systematically within our framework.
Also, the rigorous analytic proof of the N=4 Kac determinant formulae
can be accomplished by the use of the whole sets of the
charge-screening operators.
The preliminary results are already reported in ref.{\mat}.

We also mention here that
the extention of our approach\refmark{\mat}
to the N=4 SU(2)$\times$
SU(2) extended SCAs has been attempted in ref.{\ito}.
However, the construction of the representation blocks
of the N=4 SU(2)$\times$SU(2) SCAs
as in the context of the present paper
has not been  achieved yet.
It should be worth studying for making further systematic
investigations of the representation theories
of the N=4 SCAs.

These topics and other applications will be treated in our
future publications.


\vskip 2mm

\centerline{\bf\twelvecp Acknowledgements}

We thank  T. Eguchi and M. Kato at University of Tokyo and T. Uematsu
at Kyoto University
for useful discussions.
We also express our sincere thanks to Theory Division at CERN and
Theory Group at SLAC for the kind hospitality during my visit.
Most of my elaborate calculations in the present paper
were performed and
completed in a pleasant and encouraging atmosphere during my stay
at Geneva and at Stanford in 1993.

\endpage

\refout

\endpage


\centerline{\bf Figure Captions}

\vskip .5cm

\item{Fig.1}:The basic unit of the representation blocks
consisting of the sixteen conformal fields.
The eight \lq\lq boson-like\rq\rq\
and eight \lq\lq fermion-like\rq\rq\
conformal fields
(represented by circles and triangles respectively)
are assigned to the fourteen edge corners and the one
doubly degenerate center of the {\it rhombic dodecahedron} which are located
at the top and bottom points and on the three layers in between.

\item{Fig.2}:The steps of the OPE operations for $V$ to reach the top
point of $U$ are illustrated. The upper rectangles are at right angles
to those in the lower ones. Each set of conformal fields,
$\{V, V^+, \bar V_-, W^{(+)}\}$ in (a)
or
$\{V, \bar V_+, V^-, W^{(-)}\}$ in (b), forms a {\it closed}
primary multiplet representing
the corresponding $N=2$ U(1) subalgebras of the $N=4$ SU(2) SCAs.

\bye